\documentclass[aps,prd,superscriptaddress,nopreprintnumbers,nofootinbib,floatfix,showpacs]{revtex4}       
\usepackage{epsfig}
 \usepackage{graphicx}
 \usepackage{dcolumn}
\usepackage{bm}
\usepackage{subfigure}
\usepackage{amsmath, amsthm, amssymb}

\def\etap{\eta^{\prime}}
\def\ebeam{E_{\rm beam}}
\def\egg{\eta\to\gamma\gamma}
\def\ppp{\pi^+\pi^-\pi^0}
\def\gg{\gamma\gamma}
\def\eppp{\eta\to\pi^+\pi^-\pi^0}
\def\de{\Delta E}
\def\ipb{\rm pb^{-1}}
\def\mbc{M_{\rm bc}}

\def\tp{T^{\prime}}
\def\cp{C^{\prime}}
\def\ep{E^{\prime}}
\def\ap{A^{\prime}}
\def\sep{SE^{\prime}}

\begin{document}


\title{Measurement of exclusive $D$ meson decays to $\eta$ and $\etap$ final states
and SU(3) amplitude analysis}

\author{M.~Artuso}
\author{S.~Blusk}
\author{S.~Khalil}
\author{J.~Li}
\author{R.~Mountain}
\author{S.~Nisar}
\author{K.~Randrianarivony}
\author{N.~Sultana}
\author{T.~Skwarnicki}
\author{S.~Stone}
\author{J.~C.~Wang}
\author{L.~M.~Zhang}
\affiliation{Syracuse University, Syracuse, New York 13244, USA}
% ...

\collaboration{CLEO Collaboration}
\noaffiliation

\date{\today}


\begin{abstract} 
  Using 281~$\ipb$ of data collected with the CLEO-c detector, we present new measurements of Cabibbo-suppressed 
decays of $D^0$ and $D^+$ mesons to $\eta$ and $\etap$ final states. We make first observations
of $D^0\to\etap\pi^0$, $\eta\eta$, $\eta\etap$, and $\eta\pi^+\pi^-$, and find evidence for
$D^+\to\eta\pi^+\pi^0$, $D^+\to\etap\pi^+\pi^0$ and $D^0\to\etap\pi^+\pi^-$. We also
report on improved measurements of $D^0\to\eta\pi^0$, $D^+\to\eta\pi^+$ and $D^+\to\etap\pi^+$.
Using the measured two-body Cabibbo-suppressed decays, we extract amplitudes for specific
flavor topologies and compare them to those from Cabibbo-favored decays. 
\end{abstract}

\maketitle
\setcounter{footnote}{0}

\begin{figure*}[htb]
   \centering{
        \subfigure[~Tree (T)]{
        \includegraphics[scale=0.39]{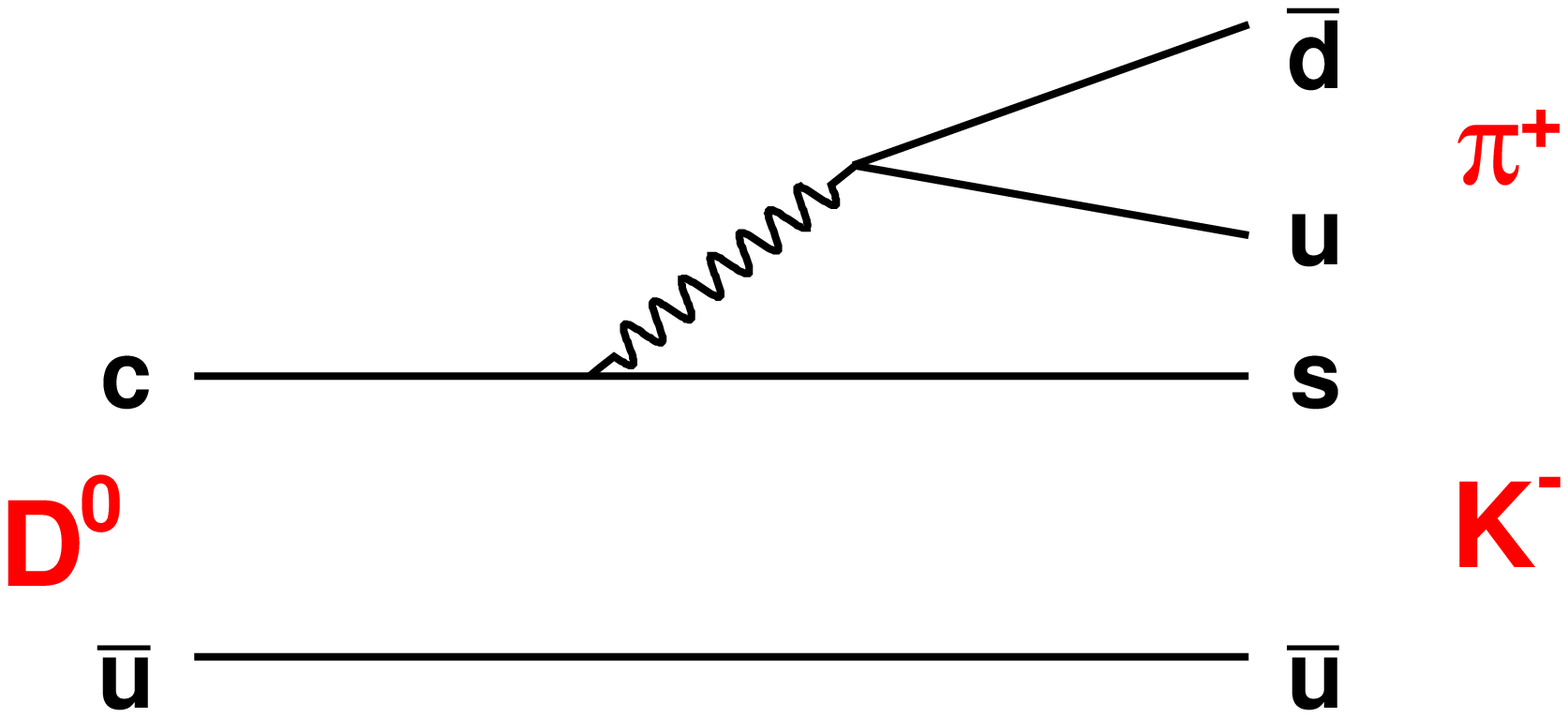}
        \label{fig:fig1}
        }
	\hfill
        \subfigure[~Color-suppressed tree (C).]{
	\includegraphics[scale=0.39]{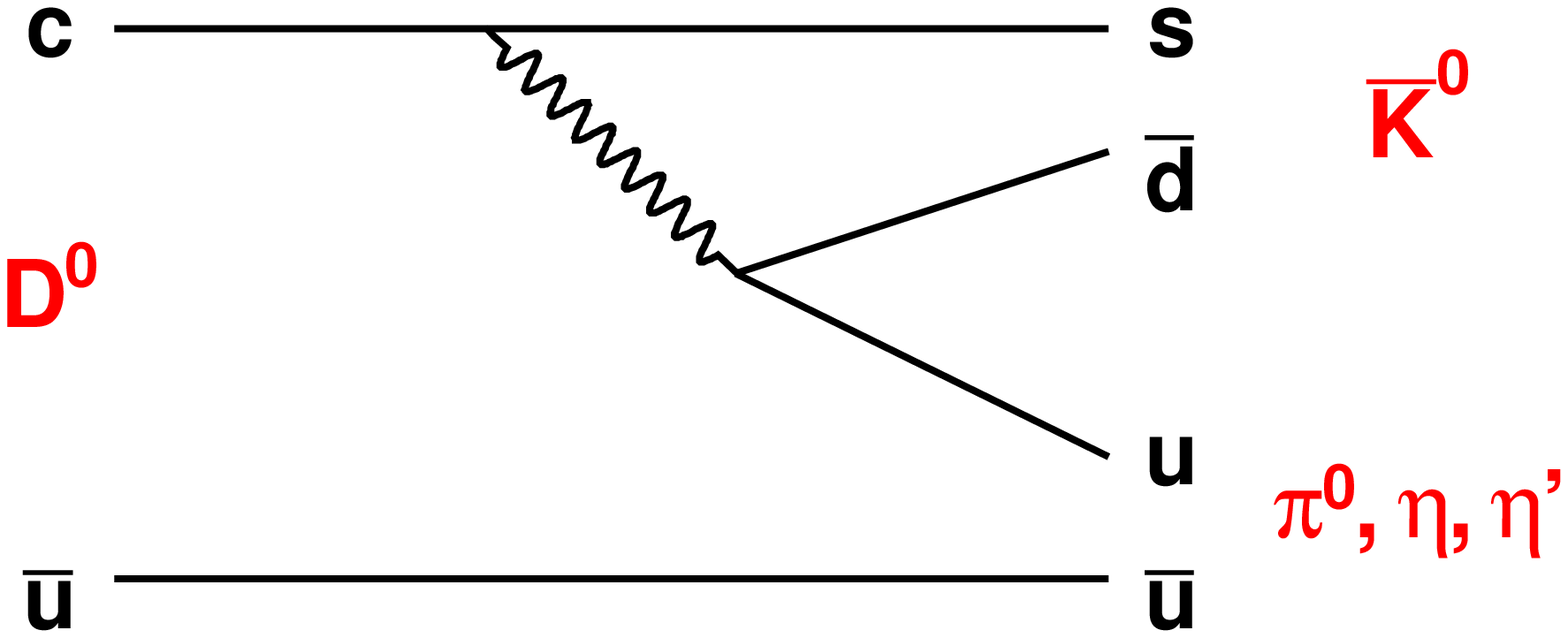}
        \label{fig:fig2}
	}		
        \subfigure[~Exchange (E).]{
	\includegraphics[scale=0.39]{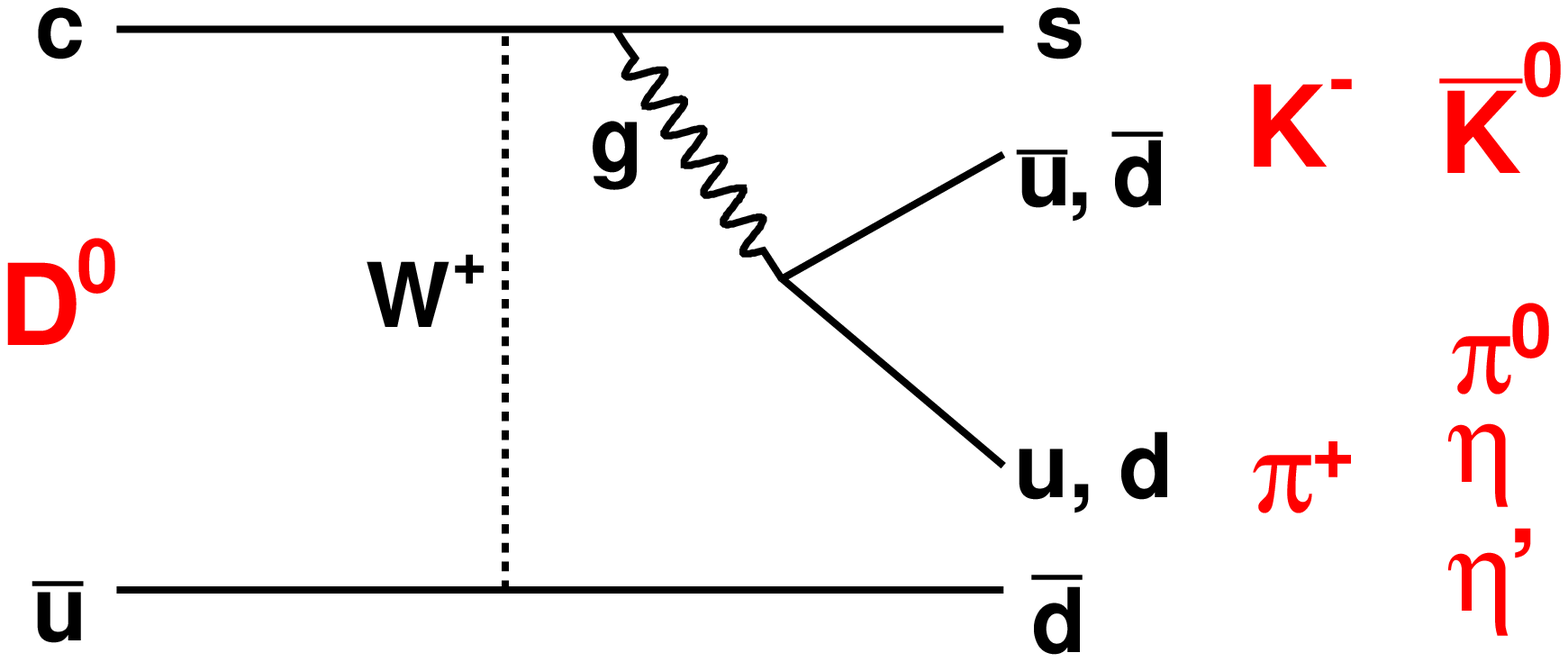}
        \label{fig:fig3}
	}		
	\hfill
        \subfigure[~Annihilation (A).]{
	\includegraphics[scale=0.39]{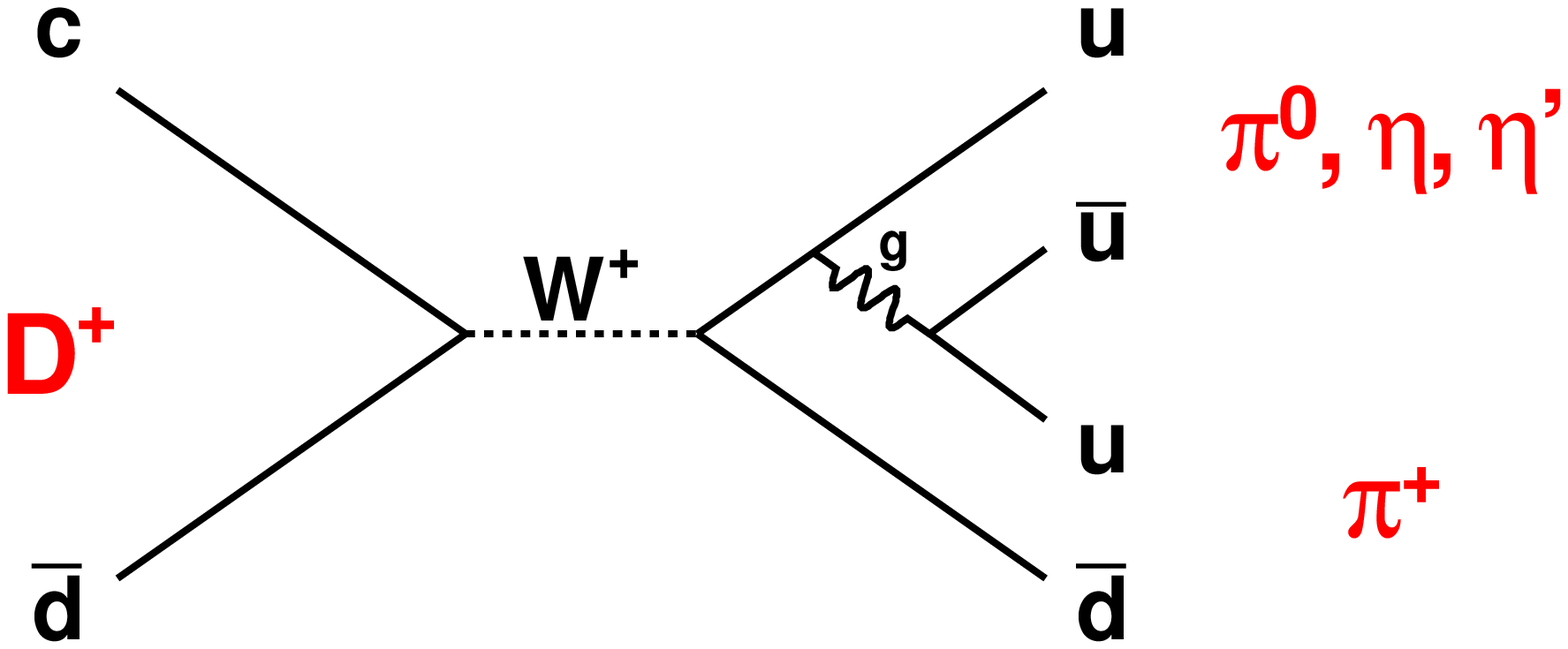}
        \label{fig:fig4}
	}		
        \subfigure[~Singlet Exchange (SE).]{
	\includegraphics[scale=0.42]{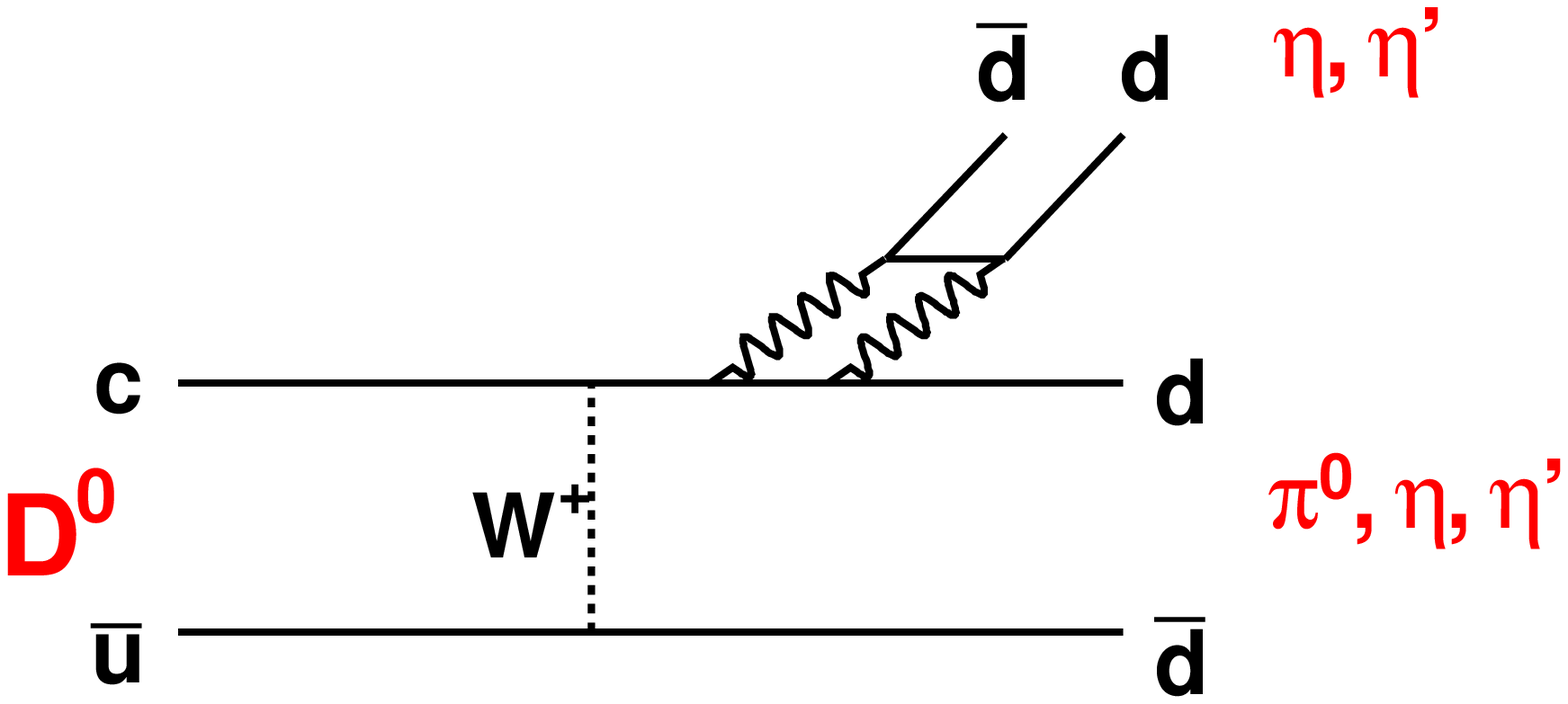}
        \label{fig:fig5}}
	}		
        \caption{Feynman diagrams for various flavor topologies used to describe $D\to PP$ decays.
	Here, we use Cabibbo-favored diagrams as an example, for $T$, $C$, $E$, and 
        Cabibbo-suppressed decays for the $A$ and $SE$ diagrams.}
   \label{fig:feynman}
\end{figure*}

      Charm decays provide a laboratory for the study of the weak and strong interactions.
Because of their simplicity, two-body hadronic decays provide experimental input for
understanding strong final state interaction effects in heavy meson 
decays~\cite{lai,guo,ablikim,rosner1,buccella,kamal}. By studying
these decays, one can determine the magnitude and relative phases of contributing 
isospin or topological decay amplitudes. To describe these decays,
particularly for $D$ and $B$ mesons, one often invokes symmetries, such as SU(3) flavor symmetry, 
which assumes the masses
of the $u$, $d$ and $s$ quarks are equal (and implicitly assumed to be small). Small differences
in decay rates are then attributed to SU(3) symmetry-breaking effects. Typically, such
symmetry-breaking effects are at the level of 20\% for $D$ meson decays, and are expected to 
decrease as the parent meson mass increases, {\it e.g.,} $B$ decays. In the SU(3) flavor
quark diagrammatic approach~\cite{rosner1}, two-body decays are
decomposed into contributions from tree (T), color-suppressed (C), annihilation (A) and 
exchange (E) diagrams.
Penguin diagrams are expected to be negligible in charm decays and are not discussed any further.
Additional disconnected diagrams, such as the singlet-exchange (SE) and singlet-annihilation (SA)
may also contribute to final states that include $\eta$ and $\etap$ via their coupling to
the SU(3) singlet component of these mesons. These quark diagrams are shown in
Fig.~\ref{fig:feynman}.


	An analysis of $D\to PP$ Cabibbo-favored (CF) decays has been carried out~\cite{rosner2},
and it was shown that their branching fractions are well-described within the SU(3) flavor 
topology approach.
In this report, we measure branching fractions of singly Cabibbo-suppressed (CS) decays and 
analyze them within the same framework~\cite{rosner3}. 

This analysis utilizes $281~\ipb$ of data collected on the $\psi(3770)$ resonance at the
Cornell Electron Storage Ring. This energy is just above threshold for
production of $D\bar{D}$ pairs, and thus no additional particles accompany the pair.
The decay products of the $D$ mesons are reconstructed using the CLEO-c detector, which
is a general purpose solenoidal detector. The detector includes a tracking system for
measuring momenta and specific ionization ($dE/dx$) of charged particles, a Ring Imaging Cherenkov
Counter (RICH) for particle identification, and a CsI calorimeter (CC) for detection of 
electromagnetic showers. The CLEO-c detector is described in detail elsewhere~\cite{cleo3}.

We reconstruct $D$ meson candidates in the following singly Cabibbo-suppressed decay 
modes: $D^+\to\eta\pi^+$, $\etap\pi^+$, $\eta\pi^+\pi^0$, $\etap\pi^+\pi^0$, and 
$D^0\to\eta\pi^0$, $\etap\pi^0$, $\eta\eta$, $\eta\etap$, $\eta\pi^+\pi^-$ and $\etap\pi^+\pi^-$.
Unless otherwise noted, charge conjugate final states are assumed throughout.

Charged particles are reconstructed using the tracking system 
and are required to pass a set of standard selection criteria~\cite{dhad}.
Charged pions are identified using a $\chi^2$ discriminant based on $dE/dx$ and RICH 
information. First, we require that the measured $dE/dx$ is within three standard deviations ($\sigma$)
of the expected value for a pion at the given momentum. We then define:

\begin{equation}
\Delta\chi^2 \equiv \chi_K^2 - \chi_{\pi}^2 + LL_K - LL_{\pi}, 
\end{equation}

\noindent where $\chi_{K(\pi)}$ is the difference in the measured and expected $dE/dx$ for the
kaon (pion) hypothesis, normalized by its uncertainty,
and $LL_{K({\pi})}$ is the negative log-likelihood for the kaon (pion) hypothesis obtained
from the RICH. 
The RICH information is used only for charged particles with momentum above 0.7~GeV/$c$
and possessing at least three associated Cherenkov photons.
We require $\Delta\chi^2>0$, which results in an efficiency of about 95\% and
a fake rate of no more than 2\%.
We construct $\pi^0\to\gamma\gamma$ and $\eta\to\gamma\gamma$ candidates
using reconstructed showers in the calorimeter that have 
energy $E_{\gamma}>30$~MeV ($E_{\gamma}>50$~MeV for $\egg$) and
no charged track that is projected within the vicinity of the shower.
For each candidate, we require the mass pull, 
$\sigma_M\equiv (M_{\rm rec}-M_{P})/\sigma_{M_{\rm rec}} < 3$, where 
 $M_{\rm rec}$ is the reconstructed invariant mass, $M_{P}$ is the parent particle mass 
(here $\pi^0$ or $\eta$), and $\sigma_{M_{\rm rec}}$ is the uncertainty in the 
invariant mass of the candidate
($\approx$5-7~MeV/$c^2$). Candidate $\eta\to\pi^+\pi^-\pi^0$ decays are also formed using
the previously discussed selection criteria on charged and neutral pions.
To improve the resolution on the $\pi^0$ or $\eta$ momentum,
the decays are kinematically constrained to the known $\pi^0$ and $\eta$ 
meson masses. The improved four-momenta are then used in subsequent analysis.
We select $\etap\to\eta\pi^+\pi^-$ candidates by requiring the mass difference, 
$402 < M_{\eta\pi^+\pi^-}-M_{\eta} < 418$~MeV/$c^2$.

For $D$ decay modes with only a single $\eta$, there is minimal benefit to including the 
$\eta\to\pi^+\pi^-\pi^0$ mode because of the lower branching fraction and detection efficiency.
We therefore only use $\eta\to\gamma\gamma$. However, for $D^0\to\eta\eta$ and $D^0\to\eta\etap$,
there is a significant gain in statistical power by allowing one $\eta\to\pi^+\pi^-\pi^0$ decay in the final state.
To reduce combinatoric background in three-body decays we veto $\eta\to\gamma\gamma$ 
candidates that share a photon with any $\pi^0$ with $|\sigma_M|<3.0$.
To ameliorate backgrounds from Cabibbo-favored modes containing a $K^0_S$ we require oppositely
charged pion pairs in the final state to be outside the invariant mass interval from 475-520 MeV/$c^2$.

     Signal $D$ candidates are formed and required to have an energy, $E_D$, consistent
with the beam energy, $\ebeam$, by requiring $\de\equiv E_{\rm D}-\ebeam$ is consistent with zero.
The mode-by-mode $\de$ selection requirements are determined from Monte Carlo simulation,
and are shown in Table~\ref{tab:sel_req}. 
Efficiencies for these decays after all analysis requirements are determined from Monte 
Carlo simulation of these decays~\cite{evtgen,geant,photos} and are also shown 
in Table~\ref{tab:sel_req}. 

\begin{table*}[hbt] 
\begin{center}  
\caption{Summary of mode-dependent $\de$ selection requirements for modes under consideration. 
Also shown are the efficiencies from signal Monte Carlo simulation. For $D^0\to\eta\eta$ and
$D^0\to\eta\etap$, we indicate the decay modes of the $\eta$'s in parentheses.
The quoted uncertainties are statistical and systematic, respectively.
\label{tab:sel_req}} 
\begin{tabular}{lcc}\hline\hline
Mode         & $\de$ Range (MeV) &~~Efficiency (\%)~~\\
\hline
$D^+\to\eta\pi^+$         &   [-28, 25]   & $46.7\pm0.5\pm1.7$     \\
$D^+\to\etap\pi^+$        &   [-19, 18]   & $27.7\pm0.5\pm1.1$     \\
$D^0\to\eta\pi^0$         &   [-45, 34]   & $30.6\pm0.5\pm1.3$      \\
$D^0\to\etap\pi^0$        &   [-38, 32]   & $17.1\pm0.4\pm0.7$       \\
$D^0\to\eta\eta$ ($\gg)(\gg$)   & [-33, 30] & $28.9\pm0.4\pm2.2$       \\
$D^0\to\eta\eta$ ($\gg)(\ppp$) & [-26, 25] & $16.5\pm0.4\pm0.7$      \\
$D^0\to\eta\etap$  ($\gg)(\gg$) & [-27, 23] & $15.9\pm0.4\pm1.2$      \\
$D^0\to\eta\etap$  ($\gg)(\ppp$)& [-23, 20] & $8.4\pm0.3\pm0.4$      \\
$D^0\to\eta\pi^+\pi^-$    &   [-22, 19]   & $29.1\pm0.5\pm1.1$       \\
$D^+\to\eta\pi^+\pi^0$    &   [-29, 24]   & $16.7\pm0.4\pm0.7$      \\
$D^0\to\etap\pi^+\pi^-$   &   [-19, 17]   & $13.3\pm0.3\pm0.5$        \\
$D^+\to\etap\pi^+\pi^0$   &   [-22, 20]   & $7.3\pm0.3\pm0.3$      \\
\hline\hline
\end{tabular} 
\end{center} 
\end{table*}

\begin{figure*}
  \includegraphics[height=0.65\textheight]{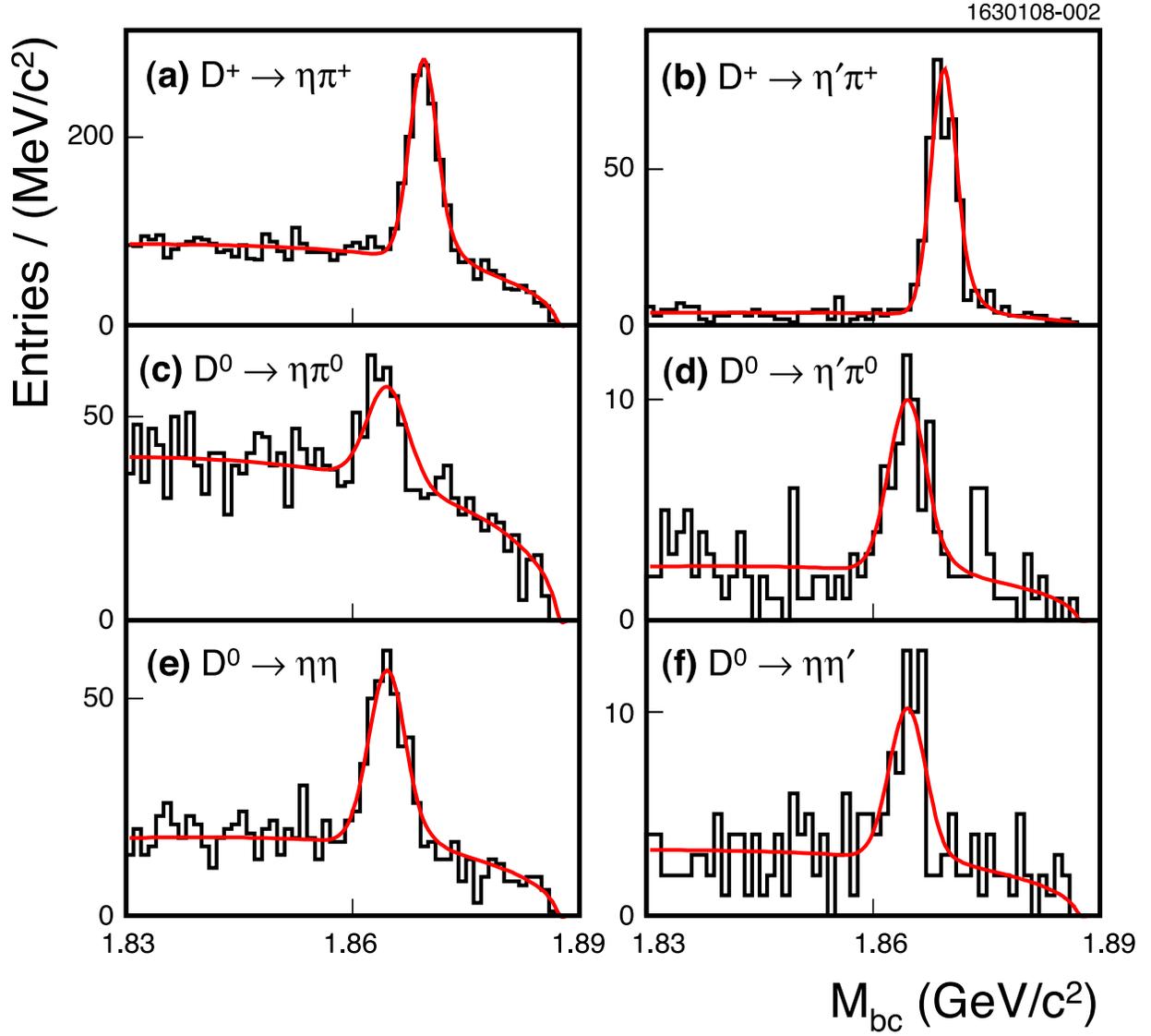}
  \caption{Distribution of $\mbc$ for the two-body Cabibbo-suppressed 
    decay modes:
    (a) $D^+\to\eta\pi^+$, (b)  $D^+\to\etap\pi^+$, (c)  $D^0\to\eta\pi^0$,
    (d) $D^0\to\etap\pi^0$, (e) $D^0\to\eta\eta$, and (f) $D^0\to\eta\etap$.
    The superimposed curve is a fit to the data as described in the text.
    \label{fig:mbc_cs_2body}}
\end{figure*}

For candidates passing these selection criteria, we compute the 
beam-constrained mass, $\mbc=\sqrt{\ebeam^2-|{\mathbf p}_D|^2}$, where ${\mathbf p}_D$ 
is the momentum of
the $D$ candidate. Substituting $\ebeam$ for the candidate energy improves the
mass resolution by about a factor of two. Figure~\ref{fig:mbc_cs_2body} shows
the $\mbc$ distributions for the two-body singly Cabibbo-suppressed decays
(a)~$D^+\to\eta\pi^+$, (b)~$D^0\to\eta\pi^0$, (c)~$D^+\to\etap\pi^+$, and 
(d)~$D^0\to\etap\pi^0$, (e)~$D^0\to\eta\eta$, and (f)~$D^0\to\etap\eta$.
Prominent peaks are observed for all six decay modes.
The points show the signal candidates in data and the
curves are fits based on maximizing the likelihood of the probability density
function, which is given by the sum of an 
{\sc argus} threshold function~\cite{argus} and an
asymmetric Gaussian signal shape ({\sc cbal})~\cite{cbal}. The {\sc argus} shape parameters 
are extracted by fitting $\mbc$ distributions obtained from the $\de$ sideband regions in data.
The high-mass tail in the $\mbc$ spectrum results from initial state radiation (ISR)
and is modeled by a power-law tail in the {\sc cbal} line-shape.
The signal shape parameters (mean, width and tail parameters) are determined from,
and fixed to the values obtained from Monte Carlo simulation.
We obtain yields of $1033\pm42$ for $D^+\to\eta\pi^+$, $160\pm24$ for $D^0\to\eta\pi^0$,
$352\pm20$ for $D^+\to\etap\pi^+$, $50\pm9$ for $D^0\to\etap\pi^0$, 
$255\pm22$ for $D^0\to\eta\eta$, and $46\pm9$ events for $D^0\to\etap\eta$.
These are first observations of $D^0\to\etap\pi^0$, $D^0\to\eta\eta$ and 
$D^0\to\etap\eta$, and correspond to signal significances of 7.2, 14.8. and 6.7, respectively. 
The significances are obtained from the differences in log-likelihood values 
with and without (signal yield set to zero) the signal component {\it{\i.e.}} $\sqrt{2(\Delta\log{L})}$ .
The yields are shown in Table~\ref{tab:yields}. For $D^0\to\eta\eta$ and $D^0\to\eta\etap$, 
we also show yields when the two $\eta$ decay channels are analyzed independently;
both are consistent with fits to the sum of both decay channels.

        Peaking backgrounds from non-resonant final states, such as $D^0\to\eta\pi^+\pi^-\pi^0$,
$D^0\to\eta\eta\pi^+\pi^-$, etc, which may contribute to the $D^0\to\eta\eta$ and $D^0\to\eta\etap$
decays, respectively, are highly suppressed because the $\pi^+\pi^-\pi^0$ and $\eta\pi^+\pi^-$
invariant masses must be consistent with the $\eta$ and $\etap$ masses, respectively.
Moreover, these
backgrounds are also Cabibbo suppressed, and thus we expect them to be very small or negligible.
This supposition is checked by selecting candidates from 
the sideband regions, $7<|\sigma_M|<10$ and $8<M_{\pi^+\pi^-\eta}-M_{\eta}-410<16$~MeV/$c^2$,
and repeating the analysis. No evidence of peaking backgrounds are found.
	
\begin{figure*}
  \includegraphics[height=0.65\textheight]{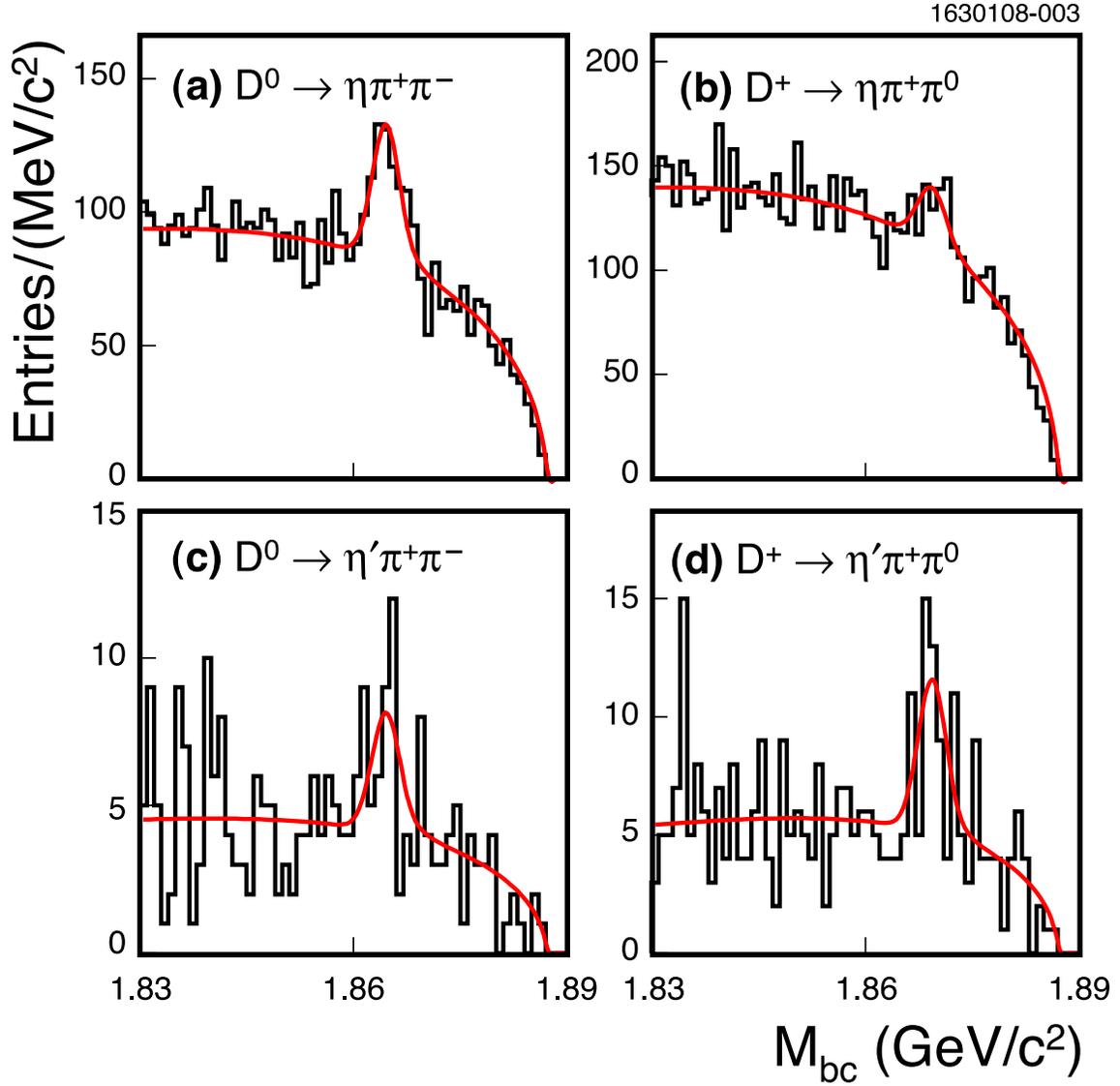}
  \caption{Distribution of $\mbc$ for the three-body Cabibbo-suppressed 
    decay modes:
    (a) $D^0\to\eta\pi^+\pi^-$,  (b) $D^+\to\eta\pi^+\pi^0$,
    (c) $D^0\to\etap\pi^+\pi^-$, and (d) $D^+\to\etap\pi^+\pi^0$.
    The superimposed curve is a fit to the data as described in the text.
    \label{fig:mbc_cs_3body}}
\end{figure*}

	Figure~\ref{fig:mbc_cs_3body} shows the $\mbc$ distributions for the
three-body Cabibbo-suppressed decays.
(a)~$D^0\to\eta\pi^+\pi^-$, (b)~$D^+\to\eta\pi^+\pi^0$, (c)~$D^0\to\etap\pi^+\pi^-$, and 
(d)~$D^+\to\etap\pi^+\pi^0$. We obtain signal yields (significances) of $258\pm32$ (9.0)
for  $D^0\to\eta\pi^+\pi^-$, $147\pm34$ (4.5) for $D^+\to\eta\pi^+\pi^0$,
$21\pm8$ (3.2) for $D^0\to\etap\pi^+\pi^-$, and $33\pm9$ (4.2) 
for $D^+\to\etap\pi^+\pi^0$, respectively. We thus establish
the $D^0\to\eta\pi^+\pi^-$ decay, and provide first evidence for the other three-body
decay modes.

The branching fractions are computed from:

\begin{equation}
{\cal{B}} = {N_{\rm sig}\over 2N_{D\bar{D}}A},
\end{equation}

\noindent where $N_{\rm sig}$ is the number of signal events, $N_{D\bar{D}}$ is
the number of $D\bar{D}$ pairs produced, and the acceptance $A=\sum{\epsilon_i {\cal{B}}_{\eta}^i}$.
Here, the sum is over the product of efficiency ($\epsilon_i$) and $\eta^{(\prime)}$ submode
branching fractions, for the measurements indicated in 
Table~\ref{tab:yields}. 
The numbers, $N_{D^0\bar{D}^0}=(1.031\pm0.015)\times10^6$ and
$N_{D^+D^-}=(0.819\pm0.013)\times10^6$, are determined from an independent measurement
of Cabibbo-favored $D$ hadronic branching fractions~\cite{dhad}. 
The yields and branching fractions are tabulated in Table~\ref{tab:yields}. For $D^0\to\eta\eta$
and $D^0\to\eta\etap$ we also show that the two decay modes yield consistent branching fractions.

     The inclusive rates for $D\to\eta^{(\prime)} X$ have been recently measured~\cite{rsia},
and the exclusive $D^0\to\eta X$, $D^0\to\etap X$ and $D^+\to\eta X$ modes measured here 
comprise about 10\% of their respective total inclusive rates. In contrast, 
${\cal{B}}(D^+\to\etap\pi^+)+{\cal{B}}(D^+\to\etap\pi^+\pi^0)\simeq~0.6\%$ accounts for
about 60\% of the total inclusive rate.

\begin{table*}[hbt] 
\begin{center}  
\caption{Summary of yields and branching fraction measurements, as discussed in
the text. For $D^0\to\eta\eta$ and $D^0\to\eta\etap$, we also show the individual results
obtained from the two $\eta$ submodes.
The first uncertainty is statistical and the second is systematic. 
Where measurements are available, results are compared to the PDG.
\label{tab:yields}}
\begin{tabular}{lccc}\hline\hline 
Mode                      & Yield      & Branching Fraction  & PDG\cite{pdg}\\
                          &            &  ($10^{-4}$)        & ($10^{-4}$) \\
\hline
$D^+\to\eta\pi^+$         &~~$1033\pm42$~~&  $34.3\pm1.4\pm1.7$ & $35.0\pm3.2$ \\        
$D^+\to\etap\pi^+$        & $352\pm20$  &  $44.2\pm2.5\pm2.9$ & $53\pm11$    \\
$D^0\to\eta\pi^0$         & $156\pm24$  &  $6.4\pm1.0\pm0.4$  & $5.6\pm1.4$ \\
$D^0\to\etap\pi^0$        & $50\pm9$    &  $8.1\pm1.5\pm0.6$  &  -\\
$D^0\to\eta\eta$          & $255\pm22$ & $16.7\pm1.4\pm1.3$ & - \\  
~~~~~($\gg)(\gg$)           & $141\pm17$ & $15.3\pm1.8$(stat.) & - \\  
~~~~~($\gg)(\ppp$)          & $115\pm13$ & $19.0\pm2.2$(stat.) & - \\  
$D^0\to\eta\etap$         & $46\pm9$    & $12.6\pm2.5\pm1.1$ & - \\
~~~~~($\gg)(\gg$)           & $33\pm8$    & $14.8\pm3.3$(stat.) & - \\
~~~~~($\gg)(\ppp$)          & $14\pm5$    & $10.5\pm3.5$(stat.) & - \\
$D^0\to\eta\pi^+\pi^-$    & $257\pm32$  & $10.9\pm1.3\pm0.9$ & $<19$ \\
$D^+\to\eta\pi^+\pi^0$    & $149\pm34$  &  $13.8\pm3.1\pm1.6$ & -   \\ 
$D^0\to\etap\pi^+\pi^-$   & $21\pm8$    & $4.5\pm1.6\pm0.5$  & - \\
$D^+\to\etap\pi^+\pi^0$   & $33\pm9$    &  $15.7\pm4.3\pm2.5$ & -   \\
\hline\hline
\end{tabular} 
\end{center} 
\end{table*}

A number of systematic uncertainties have been considered. Efficiencies, as determined
from Monte Carlo simulation, are subject to uncertainties due to finite statistics (1-2\%),
modeling of the underlying physics, and modeling of the detector response. The underlying 
physics that induces uncertainty in the efficiencies includes modeling of final state radiation (1\%)
and resonant substructure. The latter is applicable only to the three-body decay modes
and is determined by comparing efficiencies determined using a phase-space decay with 
those obtained using intermediate resonances, such as $\eta^{(\prime)}\rho$, $a_0(980)\pi$,
or $a_0(1450)\pi$. We take the largest fractional difference in efficiency as the associated 
uncertainty. The values range from 3\% for $D^0\to\eta\pi^+\pi^-$ to 12\% 
for $D^+\to\etap\pi^+\pi^0$.  Charged and neutral particle
reconstruction and identification has been extensively studied using a missing-mass 
technique~\cite{dhad}, and we find that the efficiencies in data are consistent with, or
slightly lower than simulated efficiencies.
We thus correct the simulated efficiencies as follows. 
The pion tracking efficiency and particle identification corrections are $1.000\pm0.003$
and $0.9950\pm0.0025$, respectively. For each $\pi^0\to\gamma\gamma$
($\eta\to\gamma\gamma$), we adjust the efficiencies by $0.961\pm0.020$ ($0.943\pm0.035$).
These corrections and uncertainties are included in the efficiencies shown in Table~\ref{tab:sel_req}.

Additional sources of uncertainty arise from the candidate selection requirements, namely
the $K^0_S$ veto,  $\eta^{\prime}$ mass window, $\egg$ veto (for photons also used in a $\pi^0$), 
and the restricted $\de$ range. We estimate the uncertainty introduced from these 
requirements by increasing the window size and taking the fractional difference in the
efficiency-corrected yield between the nominal and the larger window size.
For the $K^0_S$ rejection, we increase the veto region by $\pm$50\%, from which we find
an uncertainty of 2\%. For the $\eta^{\prime}$ selection, we broaden the selection
window to cover the range from 395-423~MeV/$c^2$, and find an associated uncertainty of 1\%.
The uncertainty introduced from the $\egg$ veto is only applicable to three-body decay modes.
It is estimated by comparing (data and MC simulation) the fraction of signal events that pass 
the veto with respect to no veto in the two-body decay modes. The ratio, averaged over several
modes, is $1.01\pm0.02$, and we conservatively assign a 5\% systematic error to this
source. The $\de$ systematic uncertainty is obtained by increasing the $\de$ signal window
by one unit of the Gaussian width. The corresponding uncertainty ranges from
2\% for $D^+\to\eta\pi^+$ to 4\% for $D^0\to\eta\pi^0$. Uncertainty in the signal yield receives
contributions from both the signal and background shape parameterizations. Both are studied by
varying the shape parameters one at a time by $\pm$ 1 standard deviation and adding the
resulting changes in yield in quadrature. The associated uncertainties range from
1\% (for $D\to\eta\pi^+$) to 4\% (for $D^0\to\etap\pi^+\pi^-$) for the signal shape 
uncertainty and 0.5\% (for $D\to\etap\pi^+$) to 4.0\% (for $D^0\to\etap\pi^+\pi^-$) for the 
background shape and normalization. Uncertainty due to multiple candidates in an
event is quantified by computing the average number of candidates per event within 
$\pm$5~MeV of the known $D$ mass for both simulation and data. We find good agreement
between data and simulation, except for final states with low momentum $\pi^0$'s,
where the data has a slightly larger rate of multiple candidates.
We take half the difference between the average number of multiple candidates between
data and simulation as a correction to the branching fraction, and assign 100\% uncertainty 
to it, leading to the following corrections: $1.013\pm0.013$ for 
$D^+\to\eta\pi^+\pi^0$, $1.015\pm0.015$ for $D^+\to\etap\pi^+\pi^0$, and
$1.03\pm0.03$ for $D^0\to(\egg)(\eppp)$. The rest of the modes are consistent
with unity at the level of 1\%, which we assign as a systematic
uncertainty. The branching fractions for $\egg$, $\eppp$ and $\etap\to\eta\pi^+\pi^-$ are uncertain
by 0.6\%, 1.5\% and 3.1\%, respectively, for each such decay in the final state.
Lastly, the number of $D^0\bar{D}^0$ ($D^+D^-$) events in our data sample has an 
uncertainty of 1.5\% (1.6\%)~\cite{dhad}. The systematic uncertainties for all modes under 
consideration are summarized in Tables~\ref{tab:sys} and are included in the
branching fraction measurements in Table~\ref{tab:yields}.

\begin{table*}[hbt] 
\begin{center}  
\caption{Systematic uncertainties for signal $D$ decay modes as discussed in the text.
The two entries for $D^0\to\eta\eta$ and $D^0\to\eta\etap$ correspond to the final states 
that are reconstructed with two $\egg$ decays, or one $\egg$ and one $\eppp$.
\label{tab:sys}} 
\begin{tabular}{lcccccccccc}\hline\hline
Source           & $\eta\pi^+$ & $\etap\pi^+$ & $\eta\pi^0$ & $\etap\pi^0$ & $\eta\eta$ & $\eta\etap$ & $\eta\pi^+\pi^0$ & $\etap\pi^+\pi^0$ & $\eta\pi^+\pi^-$ & $\etap\pi^+\pi^-$  \\
\hline
MC Statistics        & 1.0   & 1.0 & 1.0 & 1.0 & 1.0     & 1.0     & 2.0 & 2.0 & 1.0 & 2.0 \\
Particle Recon. \& ID& 3.7   & 3.9 & 4.3 & 4.3 & 7.4/4.3 & 7.5/4.5 & 4.3 & 4.4 & 3.8 &  4.0      \\
$\de$ Selection        & 2.0 & 2.0 & 4.0 & 3.0 & 3.0/2.0 & 2.0/2.0 & 4.0 & 3.0 & 3.0 &  2.0      \\
$\etap$ Selection      &  -  & 1.0 &  -  & 1.0 &   -     &  1.0    &  -  & 1.0 &  -  &  1.0      \\
Signal Shape	       & 1.0 & 3.0 & 1.0 & 3.0 &  3.0    &  2.0    & 2.0 & 3.0 & 3.0 &  4.0      \\
Background  	       & 0.6 & 0.5 & 2.7 & 3.2 &  1.4    &  3.1    & 1.6 & 4.4 & 2.7 &  3.2      \\
Multiple Candidates    & 1.0 & 1.0 & 1.0 & 1.0 & 1.0/3.0 & 1.0/1.0 & 1.3 & 1.5 & 1.0 &  1.0      \\
$\egg$ Veto	       &  -  &  -  &  -  &  -  &    -    &    -    & 5.0 & 5.0 & 5.0 &  5.0     \\
Resonant Substructure  &  -  &  -  &  -  &  -  &    -    &    -    & 8.0 &12.0 & 3.0 & 5.0     \\
Final State Rad.       & 1.0 & 1.0 & 1.0 & 1.0 &  1.0    &   1.0   & 1.0 & 1.0 & 1.0 & 1.0    \\
$K^0_S$ Veto	       &  -  &  -  &  -  &  -  &    -    &  -      & -   &  -  & 2.0 & 2.0     \\
$N_{D\bar{D}}$ 	       & 1.6 & 1.6 & 1.5 & 1.5 &  1.5    & 1.5     & 1.6 & 1.6 & 1.5 & 1.5      \\
${\cal{B}}_{(\etap)}$    & 0.6 & 3.2 & 0.6 & 3.2 &  1.2/2.1& 3.4/3.6 & 0.6 & 3.2 & 0.6 & 3.2       \\
\hline
Total 		       &4.9  & 6.1 & 6.9 & 7.6 &   7.7   &   8.6   &11.8 &15.7 & 8.6 & 11.4 \\
\hline\hline
\end{tabular} 
\end{center} 
\end{table*}

\begin{figure*}
  \includegraphics[height=0.35\textheight]{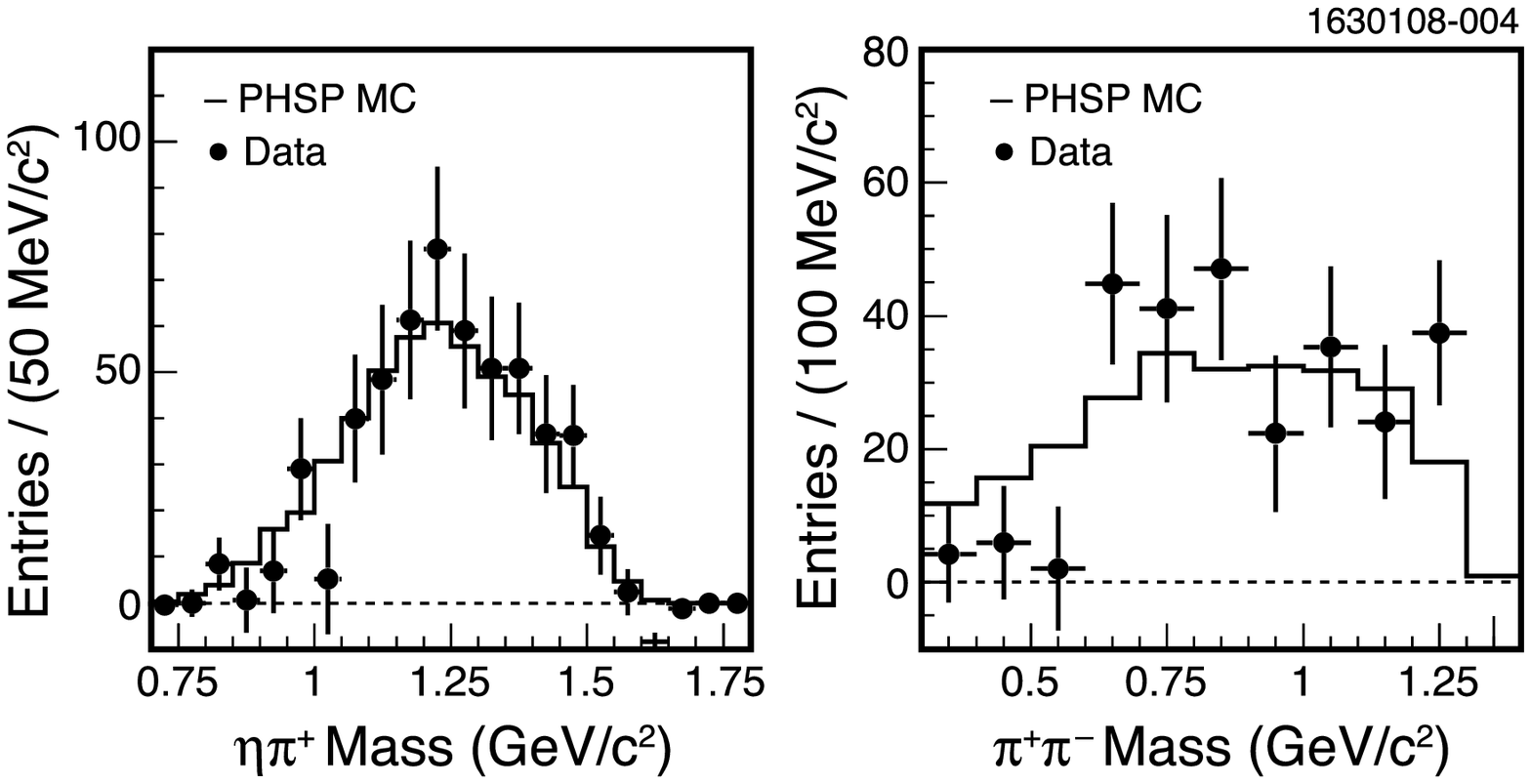}
  \caption{Sideband-subtracted invariant mass distributions for 
$\eta\pi^+$ (left) and $\pi^+\pi^-$ (right) in the
$D^0\to\eta\pi^+\pi^-$ decay. The points are data and the histogram
is a phase space (PHSP) model of the decay from Monte Carlo simulation. 
\label{fig:mpipi}}
\end{figure*}

	We have searched for intermediate resonances in $D^0\to\eta\pi^+\pi^-$. 
Figure~\ref{fig:mpipi} shows the sideband-subtracted $\eta\pi^+$ and
$\pi^+\pi^-$ invariant mass distributions. Surprisingly, there are no
significant contributions from either $\eta\rho^0$ or $a_0(980)\pi^+$.
Overlayed on the data (points) is a Monte Carlo simulation where
a phase space model is used. We find that decay is well-modeled 
by three-body phase space.

       We now apply an SU(3) diagrammatic analysis to test the validity of the SU(3) flavor
decomposition approach to charm meson decays~\cite{rosner1,rosner2,rosner3}. 
Two-body $D$ meson decays can be described using an SU(3) diagrammatic approach
in terms of an external tree diagram, $T$; 
a color-suppressed tree diagram, $C$; an exchange diagram, $E$; an annihilation
diagram, $A$; and a singlet exchange diagram, $SE$. The $SE$ contribution
represents the matrix element that produces an $\eta$ or $\etap$ through its coupling
to the SU(3) singlet portion of these mesons. Such contributions are OZI-suppressed, and 
thus expected to be small. These five diagrams are shown in Fig.~\ref{fig:feynman}.

     We first update the fitted diagrammatic amplitudes obtained from Cabibbo-favored 
decays~\cite{rosner3}. Table~\ref{tab:cfdecay} shows the SU(3) representations,
the measured branching fractions, decay momentum ($p^*$), and invariant amplitudes
for Cabibbo-favored $D\to PP$ decays.
The lifetime of the $D$ mesons and the
$D\to\bar{K}^0(\eta,\etap)$ branching fractions are taken from the PDG~\cite{pdg}, whereas
the other six branching fractions are taken from recent
CLEO measurements~\cite{dhad,dshad,dkpi}. The eight branching fraction measurements are fit
to four topological amplitudes and three relative strong phases. The phases of
$C$, $E$ and $A$ are measured with respect to $T$, which is taken to be real.
The phase, $\delta_{AB}$ is the angle subtended from amplitude B to amplitude A. 
The fitted amplitudes (in units of $10^{-6}$~GeV) and relative strong phases are found to be:

\begin{subequations}
\begin{eqnarray}
T &=& (2.78\pm0.13),\\ 
C &=& (2.04\pm0.17) e^{i(-151\pm2)^{\circ}},  \\
E &=& (1.68\pm0.12) e^{i(117\pm4)^{\circ}},  \\
A &=& (0.54\pm0.37) e^{i(-64^{+32}_{-8})^{\circ}}. 
\end{eqnarray}
\end{subequations}

\noindent The fit $\chi^2$ is 0.65 for 1 degree of freedom, indicating that the OZI-suppressed SE and SA
contributions are not needed to describe the branching fraction measurements in Cabibbo-favored decays.
These results are consistent with those obtained in Ref.~\cite{rosner3} and
the fitted branching fractions (${\cal{B}}_{\rm fit}$ in Table~\ref{tab:cfdecay}) are 
in good agreement with the experimental values.

\begin{table*}[hbt] 
\begin{center}  
\caption{Table of branching fractions, SU(3) representations~\cite{rosner3}, decay momenta, 
and invariant amplitudes for $D\to PP$ Cabibbo-favored decays. The last column shows 
the fitted branching fractions
as described in the text. The branching fractions, ${\cal{B}}_{\rm exp}$, are taken from recent 
CLEO measurements~\cite{dhad,dshad,dkpi}
when available, otherwise we use PDG values~\cite{pdg}.\label{tab:cfdecay}} 
\begin{tabular}{lccccc}\hline\hline 
Mode & Representation & ${\cal{B}}_{\rm exp}$  &~~$p^*$~~& ${\cal{A}}$           &~~${\cal{B}}_{\rm fit}$~~\\
     &                &  (\%)                  & (MeV)   &    ($10^{-6}$~GeV)  &   (\%)     \\ 
\hline
$D^0\to K^-\pi^+$    &     $T+E$                   & $3.891\pm0.077$ & 861 & $2.52\pm0.02$ & 3.899\\
$D^0\to\bar{K}^0\pi^0$ & ${1\over\sqrt{2}}(C-E)$  & $2.238\pm0.109$   & 860 & $1.91\pm0.05$ & 2.208\\
$D^0\to\bar{K}^0\eta$ & ${1\over\sqrt{3}}C$       & $0.76\pm0.11$   & 772 & $1.18\pm0.09$ & 0.76 \\
$D^0\to\bar{K}^0\etap$ & ${-1\over\sqrt{6}}(C+3E)$ & $1.87\pm0.28$   & 565 & $2.16\pm0.16$ & 1.95\\
$D^+\to\bar{K}^0\pi^+$ & $(C+T)$                  & $2.986\pm0.067$   & 862 & $1.39\pm0.02$ & 2.99\\
$D_s^+\to\bar{K}^0K^+$ & $(C+A)$                  & $2.98\pm0.17$   & 850 & $2.12\pm0.06$ & 3.02\\
$D_s^+\to\eta\pi^+$ & ${1\over\sqrt{3}}(T-2A)$     & $1.58\pm0.21$   & 902 & $1.50\pm0.10$ & 1.47\\
$D_s^+\to\etap\pi^+$ & ${2\over\sqrt{6}}(T+A)$    & $3.77\pm0.39$   & 743 & $2.55\pm0.13$ & 3.61 \\
\hline\hline
\end{tabular} 
\end{center} 
\end{table*}

We now analyze the Cabibbo-suppressed decays using the same framework.
Using our measured branching fractions for $D^0\to\eta\pi^0$, $D^0\to\etap\pi^0$, $D^0\to\eta\eta$ and
$D^0\to\eta\etap$, we extract $\cp$ and $\ep$, the analog of $C$ and $E$ for CS decays.
The topological amplitudes for these decays are given in Table~\ref{tab:csdecay1}.
We first assume $\sep=0$ and using these four measured branching fractions, we fit for 
$\cp$, $\ep$ and $\cos\delta_{\cp\ep}$.
The resulting amplitudes and relative strong phase are shown in Table~\ref{tab:csfit1}, and are 
compared to the values from
CF decays, scaled by $\lambda\equiv\tan\theta_C=0.2317$~\cite{pdg}. The values agree within about two standard deviations,
and are well within the oft cited $\sim$20\% level of SU(3) symmetry breaking effects in $D$ 
meson decays.

\begin{table*}[hbt] 
\begin{center}  
\caption{SU(3) representations~\cite{rosner3}, measured branching fractions (from this analysis),
decay momenta, and invariant amplitudes for selected Cabibbo-suppressed decays.
The branching fraction uncertainties are obtained from the quadrature sum of the
statistical and systematic uncertainties.\label{tab:csdecay1}} 
\begin{tabular}{lccccc}\hline\hline 
Mode               &~~~Representation~~~ &~~~Representation~~~& ${\cal{B}}_{\rm exp}$ & $p^*$ & ${\cal{A}}$ \\
                   &  ($\sep=0$)    & ($\sep\neq 0)$     &   ($10^{-4}$)   & (MeV) & $(10^{-7}$~GeV) \\
\hline
$D^0\to\eta\pi^0$ & ${1\over\sqrt{6}}(\cp-2\ep)$ & ${1\over\sqrt{6}}(\cp-2\ep-\sep)$   & $6.4\pm1.1$   & 846 & $3.26\pm0.28$ \\
$D^0\to\etap\pi^0$ & ${1\over\sqrt{3}}(\cp+\ep)$ & ${1\over\sqrt{3}}(\cp+\ep+2\sep)$ & $8.1\pm1.6$   & 678 & $4.09\pm0.40$ \\
$D^0\to\eta\eta$ & ${2\sqrt{2}\over 3}\cp$ & ${2\sqrt{2}\over 3}(\cp+\sep)$ & $16.7\pm1.9$   & 755 & $5.57\pm0.32$ \\
$D^0\to\eta\etap$ & ${-1\over 3\sqrt{2}}(\cp+6\ep)$ &~~~${-1\over 3\sqrt{2}}(\cp+6\ep+7\sep)$~~~&~~$12.6\pm2.7$~~& 537 & $5.74\pm0.61$ \\
\hline\hline
\end{tabular} 
\end{center} 
\end{table*}

\begin{table*}[hbt] 
\begin{center}  
\caption{Solutions for the topological amplitudes using Cabibbo-suppressed decays.
The fit results are compared to the values from Cabibbo-favored decays, scaled by 
$\lambda\equiv\tan\theta_C=0.2317$.
\label{tab:csfit1}} 
\begin{tabular}{ccccc}\hline\hline 
Amplitude & \multicolumn{2}{c}{Magnitude}  & \multicolumn{2}{c}{$\delta_{CE}$ ($^{\circ}$)}\\
            & \multicolumn{2}{c}{($10^{-7}$ GeV)}  & \multicolumn{2}{c}{} \\
\cline{2-5}
          &      CS       &   $\lambda\times$CF   &   CS     & CF \\ 
\hline
C         & ~$5.8\pm0.3$~ & ~$4.7\pm0.4$~ &  -  &     - \\
E         & ~$3.5\pm0.3$~ & ~$3.9\pm0.3$~ & ~$77\pm7$~ & ~$92\pm4$ \\
\hline\hline
\end{tabular}  
\end{center} 
\end{table*} 

We now allow for an additional singlet exchange amplitude, $\sep$~\cite{rosner3}.
The SU(3) representations, including $\sep$, are shown in Table~\ref{tab:csdecay1}. 
Invoking SU(3) symmetry, we have
$\tp=\lambda T$, $\cp=\lambda C$, $\ep=\lambda E$ and $\ap=\lambda A$,
$\delta_{\cp\tp}=\delta_{CT}$, $\delta_{\ep\tp}=\delta_{ET}$, and $\delta_{\ap\tp}=\delta_{AT}$.
The amplitudes $\tp$, $\cp$, $\ep$, and $\ap$ (in units of $10^{-7}$~GeV) are found to be:

\begin{widetext}
\begin{subequations}
\label{eq:csamp}
\begin{eqnarray}
\tp &=& 6.44\pm0.30, \\ 
\cp &=& 4.73\pm0.39 e^{i(-151\pm2)^{\circ}}=(-4.15\pm0.38)+ i (-2.25\pm0.15),\\
\ep &=& 3.89\pm0.28 e^{i(-117\pm4)^{\circ}}=(-1.76\pm0.24)+ i (3.48\pm0.29), \\
\ap &=& 1.25\pm0.86 e^{i(-64^{+32}_{-8})^{\circ}}=(0.55\pm0.34)+ i (-1.14\pm0.83). 
\end{eqnarray}
\end{subequations}
\end{widetext}

\begin{widetext}

One may rewrite the amplitudes in Table~\ref{tab:csdecay1} as follows~\cite{rosner3}:

\begin{subequations}
\begin{eqnarray}
-\sqrt{6}{\cal{A}}(D^0\to\eta\pi^0) &=& 2\ep-\cp+\sep  =(0.63+i9.21)\times10^{-7}~{\rm GeV}+\sep,  \label{eq:cs1} \\
{\sqrt{3}\over{2}}{\cal{A}}(D^0\to\etap\pi^0) &=& {1\over 2}(\cp+\ep)+\sep = (-2.95+i0.61)\times10^{-7}~{\rm GeV}+\sep  \\
{3\over{2\sqrt{2}}}{\cal{A}}(D^0\to\eta\eta) &=& \cp+\sep = (-4.15-i2.25)\times10^{-7}~{\rm GeV}+\sep \\
{-3\sqrt{2}\over{7}}{\cal{A}}(D^0\to\eta\etap) &=& {1\over 7}(\cp+6\ep)+\sep = (-2.10+i2.66)\times10^{-7}~{\rm GeV}+\sep 
\label{eq:cs4}
\end{eqnarray}
\end{subequations}
\end{widetext}

\noindent The right hand side of each of these four equations defines a vector in the complex plane,
which contains an unknown complex offset, $\sep$.
The left hand side of these equations defines four circles in the complex plane,
whose radii are determined by their measured branching fractions. We thus have
four constraints with which to solve for the real and imaginary parts of $\sep$.
The solutions are obtained graphically from the common intersection point(s) of 
these four circles.

\begin{figure*}[hbt]
  \includegraphics[height=0.75\textheight]{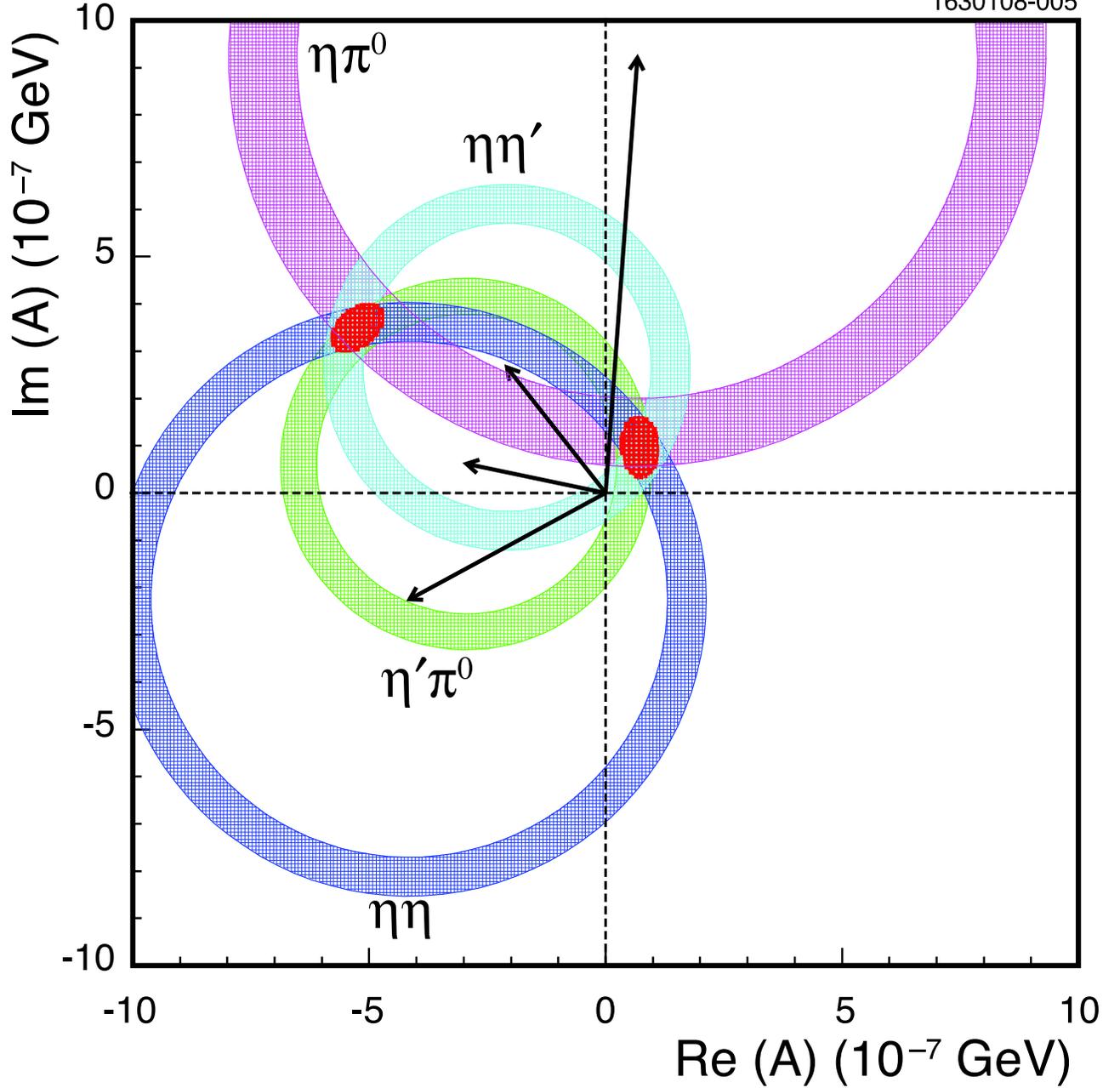}
  \caption{Graphical presentation of the four Cabibbo-suppressed amplitudes as
circles in the complex plane. The arrows correspond to their centers, as
determined from Cabibbo-favored decays, scaled by $\tan\theta_C=0.2317$. The two overlap
regions correspond to the allowed $\sep$ solutions.\label{fig:scs_amp}}
\end{figure*}

	Figure~\ref{fig:scs_amp} shows Eqs.~\ref{eq:cs1}-\ref{eq:cs4} graphically, where
we find two common intersection points, which are (the negative of) the allowed $\sep$ solutions.
The small solution Re$(\sep)=(-0.7\pm0.4)\times10^{-7}$~GeV and 
Im$(\sep)=(-1.0\pm0.6)\times10^{-7}$~GeV,
is favored due to the expected OZI suppression. The larger solution, which is
disfavored, corresponds to Re$(\sep)=(5.3\pm0.5)\times10^{-7}$~GeV and 
Im$(\sep)=(-3.5\pm0.5)\times10^{-7}$~GeV.
We thus find that apart from a small additional $\sep$ contribution
to CS decays, $\cp\simeq\lambda C$ and $\ep\simeq\lambda E$, and thus these
decays respect SU(3) symmetry at the level of $\sim$20\%. We therefore find that this
SU(3) flavor diagrammatic approach provides a reasonable description of both Cabibbo-favored 
and Cabibbo-suppressed $D\to PP$ decays that depend on $C$ and $E$ diagrams.
This SU(3) topological approach may not apply equally well to all decays. For example, 
it is well known that the prediction ${\cal{B}}(D^0\to\pi^+\pi^-)={\cal{B}}(D^0\to K^+K^-)$ is not
realized due to SU(3) symmetry-breaking effects in the form factors and decays constants.

    In summary we report on first observations of $D^0\to\etap\pi^0$, $D^0\to\eta\eta$, 
$D^0\to\eta\etap$, and $D^0\to\eta\pi^+\pi^-$. We also find evidence for the three-body decays
$D^+\to\eta\pi^+\pi^0$, $D^0\to\etap\pi^+\pi^-$ and $D^+\to\etap\pi^+\pi^0$. We have 
analyzed $D^0\to\eta\pi^0,\etap\pi^0,~\eta\eta$ and $D^0\to\eta\etap$ decays within the 
SU(3) flavor topology approach~\cite{rosner3} and find that the color-suppressed and 
exchange amplitudes have magnitudes and a
relative strong phase that are consistent with Cabibbo-favored decays. We have performed
a second fit where we allow for an additional singlet-exchange amplitude and find a solution near 
zero (favored) and a larger solution, which is disfavored due to the OZI suppression of this process.

We gratefully acknowledge the effort of the CESR staff
in providing us with excellent luminosity and running conditions.
D.~Cronin-Hennessy and A.~Ryd thank the A.P.~Sloan Foundation.
This work was supported by the National Science Foundation,
the U.S. Department of Energy, the Natural Sciences and Engineering Research Council of Canada,
and the U.K. Science and Technology Facilities Council. We also thank Jon Rosner
and Bhubanjyoti Bhattacharya for their helpful comments on this manuscript.



\end{document}